\documentclass[10pt]{cai26}

\begin{document}
\def\conferenceyear{2026}
\volumeheader{39}{0}
\begin{center}

\title{Toward Autonomous SOC Operations: End-to-End LLM Framework for Threat Detection, Query Generation, and Resolution in Security Operations}
\maketitle

\thispagestyle{empty}
\pagenumbering{gobble}

\begin{tabular}{cc}
Md Hasan Saju\upstairs{\affilone,*}, Akramul Azim\upstairs{\affilone}
\\[0.25ex]
{\small \upstairs{\affilone} Department of Electrical, Computer and Software Engineering}\\
{\small Ontario Tech University, Oshawa, Ontario, Canada}\\
\end{tabular}
  
\emails{
  \upstairs{*} mdhasan.saju@ontariotechu.ca
}
\vspace*{0.1in}
\end{center}

\begin{abstract}
Security Operations Centers (SOCs) face mounting operational challenges. These challenges come from increasing threat volumes, heterogeneous SIEM platforms, and time-consuming manual triage workflows. We present an end-to-end threat management framework that integrates ensemble-based detection, syntax-constrained query generation, and retrieval-augmented resolution support to automate critical security workflows. Our detection module evaluates both traditional machine learning classifiers and large language models (LLMs), then combines the three best-performing LLMs to create an ensemble model, achieving 82.8\% accuracy  while maintaining 0.120 false positive rate on SIEM logs. We introduce the SQM (Syntax Query Metadata) architecture for automated evidence collection. It uses platform-specific syntax constraints, metadata-based retrieval, and documentation-grounded prompting to generate executable queries for IBM QRadar and Google SecOps. SQM achieves a BLEU score of 0.384 and a ROUGE-L score of 0.731. These results are more than twice as good as the baseline LLM performance. For incident resolution and recommendation generation, we demonstrate that integrating SQM-derived evidence improves resolution code prediction accuracy from 78.3\% to 90.0\%, with an overall recommendation quality score of 8.70. In production SOC environments, our framework reduces average incident triage time from hours to under 10 minutes. This work demonstrates that domain-constrained LLM architectures with retrieval augmentation can meet the strict reliability and efficiency requirements of operational security environments at scale.

\end{abstract}

\copyrightnotice


\section{Introduction}

Security Operations Centers (SOCs) are under increasing pressure due to the rapid growth of attacks, higher alert volumes, and the complexity of modern enterprise systems. Large organizations generate millions of security events every day from network devices, endpoints, cloud services, and identity platforms. These events are aggregated by Security Information and Event Management (SIEM) systems, which form the core of SOC monitoring and investigation workflows. SIEM systems have their querying and correlation capabilities, but still rely heavily on manual analyst effort for alert interpretation, evidence gathering, and incident resolution. Vielberth et al.\cite{Vielberth} identify an insufficient level of automation as one of the most critical open challenges in SOC operations, with many investigative tasks still requiring substantial manual effort in environments where skilled human resources are scarce.

Incident investigation remains a primary bottleneck in SOC operations. Analysts must assess alert severity, construct platform specific queries to retrieve supporting evidence, correlate information across multiple data sources, and determine an appropriate resolution. This process requires substantial domain expertise, including proficiency in proprietary query languages such as AQL and YARA-L, as well as familiarity with organizational context and attack patterns. Consequently, incident investigation often takes hours to days, contributing to alert fatigue, delayed response, and increased risk of missed or improperly handled threats. Recent advances in large language models (LLMs) have shown promise in reasoning over unstructured text and assisting with complex analytical tasks. However, directly applying general purpose LLMs to SOC workflows exposes several limitations. Many existing approaches focus on isolated subtasks, such as detection or summarization, rather than supporting the full end to end investigation and resolution workflow used by analysts in production environments.

In this work, We present an end-to-end threat management framework that augments SOC analysts workflow through three tightly coupled components. First, a threat detection module that combines traditional machine learning models with multiple large language models using ensemble reasoning to identify critical security events from SIEM logs while balancing accuracy, precision, recall and false positive rate. Second, a syntax constrained evidence collection mechanism, termed {Syntax Query Metadata} (SQM), which enables reliable and executable query generation across heterogeneous SIEM platforms by grounding LLM outputs in platform specific syntax rules, metadata driven retrieval, and official documentation. Third, a threat resolution and recommendation module that leverages SQM derived evidence and historical SOC knowledge to assist analysts in selecting resolution codes, generating justifications, and recommending remediation actions.

We evaluate the proposed framework using real world SIEM logs and incident tickets from production SOC environments. Experimental results show that ensemble based LLM detection outperforms traditional machine learning baselines while reducing false positives compared to individual models. The SQM architecture improves query generation performance to more than twice that of the baseline models. It can generate executable queries for both IBM QRadar and Google SecOps. Most importantly, integrating SQM derived evidence into downstream resolution tasks improves resolution code prediction accuracy from 78.3\% to 90.0\%. 

The main contributions of this paper are as follows:
\begin{itemize}
    \item We propose an end-to-end threat management framework that integrates threat detection, evidence collection, and incident resolution into a unified SOC workflow powered by large language models and tested on real-world production data.
    \item We introduce the SQM architecture for SIEM aware query generation, combining syntax allow lists, metadata driven retrieval, and documentation grounded prompting to produce reliable and executable queries across heterogeneous platforms.
    \item We show that SQM-derived evidence, combined with Retrieval-Augmented Generation (RAG), significantly improves incident resolution accuracy and recommendation quality.
\end{itemize}

\section{Background \& Literature Review}
\subsection{Background}
A Security Operations Center (SOC) is a centralized facility where analysts monitor, detect, and respond to cybersecurity incidents. Security Information and Event Management (SIEM) platforms aggregate and correlate security event data from across an organization's infrastructure, enabling centralized querying and alerting. 

AQL (Ariel Query Language) is IBM QRadar's proprietary query language for searching event and flow data, while YARA-L 2.0 is Google SecOps' detection rule language used to define threat patterns over log events. 

Retrieval-Augmented Generation (RAG) is a technique that enhances LLM outputs by retrieving relevant external documents and injecting them into the model's prompt context at inference time.


\subsection{Literature Review}
Recent efforts to automate SOC workflows have focused on individual stages of the incident lifecycle, but no unified framework covers detection, evidence collection, resolution, and recommendation generation together. In playbook-assisted response, Akbari Gurabi et al. \cite{akbari2024requirements} conducted expert interviews and identified 16 requirements for playbook-driven incident response and reporting. Based on this, they proposed a conceptual architecture that integrates the CACAO standard with TheHive, Cortex, and MISP. Their framework is strong in requirements analysis, but it still depends on manually written playbooks and lacks automation. Paduraru et al. \cite{paduraru2025automated} addressed this limitation with CyberPlaybookLLM, a fine-tuned LLaMA 3.1-8B model that generates CACAO-compliant playbooks from incident descriptions. Their model achieved 84.3\% mitigation precision and 96.8\% schema validity. However, their work focuses only on response orchestration and does not cover upstream detection or SIEM query generation. Moving further toward full incident lifecycle automation, Ismail et al. \cite{ismail2025toward} proposed an agentic LLM-powered SOAR platform with the IVAM framework. This framework organizes investigation, validation, and active monitoring using MITRE ATT\&CK mapping and quantitative risk assessment. Their multi-agent system reduced a traditional 38-step playbook to ten key actions on a Wazuh SIEM deployment. However, the validation was limited to a virtualized environment, and the framework does not include cross-platform SIEM query generation or data-driven resolution prediction based on historical incidents.

For cross-platform SIEM query generation, Saju et al. \cite{saju2025synrag} proposed SynRAG, a RAG-based framework that creates platform-specific queries from YAML threat specifications using documentation grounding and syntax allow-lists. SynRAG achieved a BLEU score of 0.1287, a ROUGE-L score of 0.6039, and 85\% query executability across QRadar AQL and Google SecOps YARA-L 2.0. However, it works as a standalone tool and is not integrated with detection or resolution pipelines. It also does not use metadata embeddings for query generation. Ojuri et al. \cite{ojuri2025optimizing} studied LLM-based text-to-SQL conversion using ReAct agents. They showed that fine-tuned GPT-3.5-turbo can reach 97\% syntactic validity and 91.5\% execution accuracy on relational databases. Their results show that LLM-based query generation can work well with schema grounding. However, this approach does not directly transfer to SIEM query languages, because those languages need security-specific temporal operators and event correlation logic that standard SQL does not provide.

At the detection layer, Tendikov et al.\cite{tendikov2024security} built a network intrusion detection system using Azure Sentinel honeypot data augmented with CICIDS2017 logs, achieving 0.97 F1-score with a Random Forest Classifier after feature selection and hyperparameter tuning. While practical, their approach relies on manual feature engineering over network traffic features and does not leverage LLMs for semantic log analysis, query generation, or downstream resolution. Charla\cite{charla2025ai} proposed a broader AI-enhanced SIEM architecture incorporating classification, explainability (SHAP, LIME), and playbook orchestration, reporting detection rates above 93\% and a 60\% reduction in compliance reporting time across a multi-cloud testbed. However, the evaluation used synthetic attack scenarios rather than production data, and the architecture does not incorporate LLM-driven query generation or resolution recommendation.

Collectively, the existing literature addresses playbook generation\cite{akbari2024requirements}\cite{paduraru2025automated}, SIEM query synthesis\cite{saju2025synrag}, structured query optimization\cite{ojuri2025optimizing}, log classification\cite{tendikov2024security}, and explainable AI-driven triage\cite{charla2025ai} as disjoint problems. No prior work provides a unified LLM framework that provides a single pipeline that tightly couples threat detection, executable SIEM query generation for evidence collection, and resolution prediction . None of these previous approaches generate actionable recommendations tailored to the specific incident context. So, analysts have to manually interpret detection outputs and formulate response actions without automated guidance. Our work bridges this gap through an end-to-end architecture integrating an LLM ensemble for detection which has an accuracy of 82.8\% and FPR of 0.120. Followed by a novel SQM architecture for cross-platform query generation with BLEU score of 0.384, and ROUGE-L of 0.731 (88\% executable without modification), and RAG-augmented resolution that leverages SQM-derived evidence to achieve 90.0\% accuracy alongside concrete analyst-facing recommendations for incident remediation—reducing incident triage time. To the best of our knowledge, this is the first work demonstrating that LLM-generated evidence queries materially improve downstream resolution accuracy and recommendation quality, closing the loop between detection, investigation, and response.

\section{Methodology}
\label{methodology}
The proposed framework is tested using real-world production data across three closely related components. For threat detection, we use 20,000 SIEM log entries collected from different vendor environments, including network devices, endpoints, and cloud services. Using data from multiple vendors helps the model perform better across different log formats and event structures. The dataset includes 40\% critical events and 60\% non-critical events, which matches a realistic SOC alert distribution. After the detection, each log is given a composite risk score by that component. This score combines the SIEM-generated magnitude value with the model’s predicted criticality probability. It helps analysts focus on high-risk events first and investigate them more quickly. To evaluate evidence collection and incident resolution, we use real-world analyst query requirements and ServiceNow incident tickets from different offense types and scenarios. The results are validated using human-in-the-loop validation where domain experts and SOC analysts verify the logical correctness, accuracy, and effectiveness of the approach inquiring suggestions generated by the framework. The analysts' input is used to validate the corresponding automated scores and ensure that the scores accurately reflect the quality of outputs produced by the framework.

\subsection{Threat Detection}
\begin{figure*}[t]
\centering
\includegraphics[width=\textwidth]{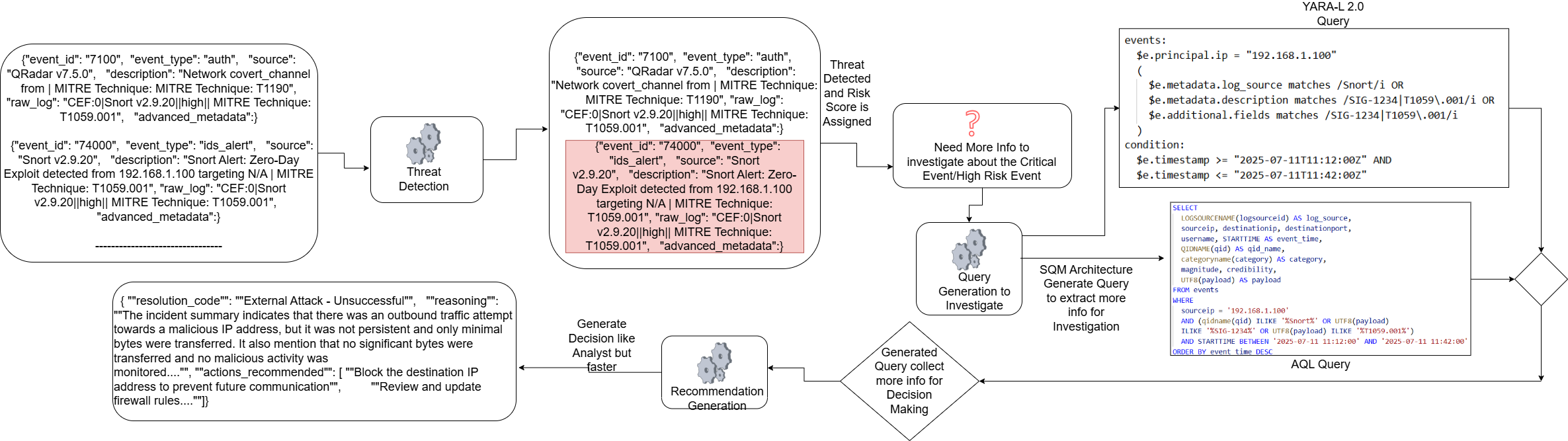}
\caption{End-to-End Threat Management Framework Overview}
\label{fig:framework_overview}
\end{figure*}

The detection component which is illustrated in Figure \ref{fig:framework_overview} identifies critical security events from SIEM logs automatically. This is set up as a supervised binary classification problem, where each log entry is labeled as either critical or non-critical based on its severity and contextual risk indicators. Before using the ML models, the dataset goes through several preprocessing steps. These include removing duplicates, filling in missing values, encoding categorical features, normalizing numerical features, and extracting temporal features from timestamps. Nested and semi-structured fields, such as metadata and behavioral analytics, are parsed and flattened to enrich the feature space. Text fields like event descriptions and log messages are also processed before using them as a feature. These are converted into dense numerical representations using the all-MiniLM-L6-v2 embedding algorithm. This allows the models to work with semantic information alongside the structured features.

To reduce dimensionality and improve discriminative performance, a hybrid feature selection strategy is employed. Statistical methods (Chi-Square and ANOVA F-tests), information-theoretic measures (Mutual Information), and model-based approaches (Random Forest feature importance and Recursive Feature Elimination) are applied independently. Feature scores from these methods are normalized and averaged to compute a unified importance ranking, from which the top features are selected for training. Using this optimized feature subset, five traditional machine learning classifiers are trained and evaluated. Model performance is assessed using accuracy, precision, recall, and F1-score, with particular emphasis on the false positive rate to reduce alert fatigue of the analysts.

In parallel, six large language models (LLMs) are evaluated for the same detection task by converting SIEM logs into structured textual prompts that encapsulate event context, metadata, and behavioral signals. To improve detection robustness and reduce individual model bias, the three best-performing LLMs are combined using a majority-voting ensemble strategy. Following ensemble-based classification, detected critical events are prioritized using a risk scoring mechanism derived from SIEM-generated metadata. Each log source includes a magnitude field assigned by the SIEM platform, and that quantifies the estimated impact of the event on the overall system. By combining this magnitude value with the predicted probability of criticality from the detection module, a composite risk score is computed for each events. This risk score enables prioritization of detected threats\cite{Khan2024Automated}. It ensures that events requiring immediate analyst intervention are investigated ahead of those that can be addressed with lower urgency. In practice, this prioritization layer transforms the detection output from a flat set of critical alerts into a ranked queue. And that allows SOC analysts to allocate investigative effort proportionally to the operational risk posed by each event. The prioritized critical events are then forwarded to the subsequent evidence collection and query generation phase of the framework.

\subsection{Evidence Collection: Query Generation}

\begin{figure*}[t]
\centering
\includegraphics[width=\textwidth, height=0.34\textheight, keepaspectratio]{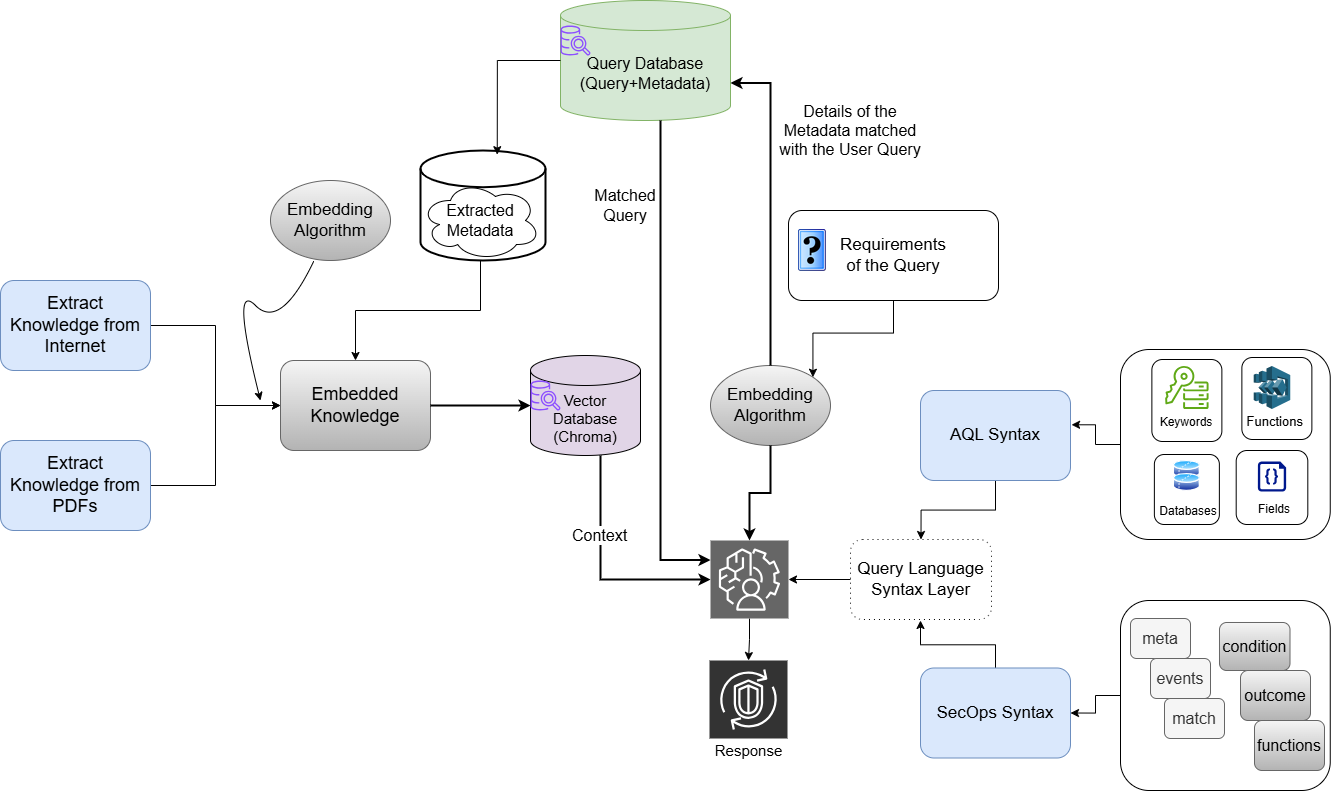}
\caption{Query Generation Architecture}
\label{fig:query_generation}
\end{figure*}

The second component of the framework which is illustrated in Figure \ref{fig:query_generation}, termed SQM (Syntax Query Metadata) architecture, addresses evidence acquisition for logs predicted as critical by the detection module. The goal is to reduce SIEM dependency on platform-specific expertise by enabling automated query retrieval and generation for heterogeneous SIEMs, specifically IBM QRadar using AQL and Google SecOps using YARA-L 2.0. SQM operationalizes query construction as a retrieval-augmented generation process constrained by SIEM-specific syntax, structured query metadata, and platform-specific query writing guidelines. In the Syntax stage, we curated platform constraints by collecting reserved keywords, functions, field names, and clause ordering rules from each SIEM. We implemented platform-specific scripts to extract and normalize these syntax elements and stored them as explicit allow-lists that act as hard constraints during query generation, ensuring that produced queries remain executable and consistent with the target SIEM grammar.

In the Query and Metadata stages, we maintained a query repository consisting of two coupled resources. First, a query database stores executable reference queries for each platform. Second, a metadata layer stores structured descriptors for each query, including query name, category, subcategory, use case, description, tags, and search keywords. To enable semantic retrieval, the metadata text is embedded using a sentence embedding model(all-MiniLM-L6-v2) and indexed in a vector database. At runtime, an analyst question, which is basically requirement of the query, is embedded and used to perform top-k nearest-neighbor search over metadata embeddings. The closest metadata entries are then used to retrieve the corresponding executable queries from the query database. These retrieved queries function as high-similarity exemplars that anchor query generation and reduce ambiguity in operator selection, field usage, and aggregation patterns.

To further improve correctness, SQM incorporates a documentation-based knowledge base built from official AQL and YARA-L 2.0 documentation, which provides platform rules and guidance for constructing valid queries. The final query generation is performed using a retrieval augmented prompting pipeline that injects three forms of context into the LLM input: (1) the SIEM syntax allow-lists, (2) documentation excerpts describing query construction rules, and (3) the top retrieved exemplar queries and their intent metadata. The model is instructed to output only the executable query, and the output is post-processed using deterministic extraction rules to remove formatting artifacts and isolate the final query string. This SQM design enables consistent, SIEM-aware evidence collection by combining syntax constraints with metadata-driven retrieval and documentation-grounded generation, thereby reducing reliance on specialized query-language expertise during SOC investigations.

\subsection{Threat Resolution and Recommendation Generation}
The final module of the proposed framework is designed to support the last stage of analyst activity, which is incident resolution and documentation. After threat detection and contextual evidence retrieval, this module treats incident closure as a structured decision support task. Given an incident ticket, the framework helps analysts by recommending a resolution code from a predefined set of closure categories used in industrial SOC environments. These categories are: Benign Positive, External Attack–Unsuccessful, External Attack–Successful, Insider Threat–Unsuccessful, Insider Threat–Successful, Escalated with No Response, and False Positive. Along with the predicted resolution code, the framework also generates a short justification for the classification and suggests appropriate response or remediation actions. This design tackles a real operational challenge. Manual resolution is often time-consuming and can vary depending on the analyst handling the case. By standardizing the process, the framework promotes consistency and helps close incidents faster.

The framework works on incident tickets sourced from the incident management platform called ServiceNow. Each ticket contains structured metadata, including offense category, severity, timestamps, source and destination attributes, flow and event statistics, and credibility indicators etc. Analyst-written resolution codes and notes from closed tickets (past tickets) are used as ground truth for evaluation. Before inference, each incident is transformed into a structured textual representation. That representation includes all relevant fields and contextual indicators. Multiple state-of-the-art large language models are evaluated through a unified execution interface using identical inference settings to ensure reproducibility. Resolution support is framed as a constrained classification task. In this classification task, the model must select from the predefined set of operational resolution codes and output a strictly formatted JSON object containing the predicted code, justification, and recommended actions.

To ground model outputs in organizational context, this module incorporates a retrieval augmented generation strategy using historical SOC knowledge. Two years of closed ServiceNow tickets, serving as historical resolution data, along with analyst runbooks, which are industry-standard procedural guides for threat handling, are collected, preprocessed, chunked, and embedded into a vector database. Recent tickets, which are part of the test set for the framework, are not included in the historical data. For each new incident, the most semantically relevant historical cases and procedural guidance are retrieved and injected into the model prompt. Evaluation is conducted under three experimental settings: resolution generation without SQM derived evidence, resolution generation with SQM query generated evidence, and an proposed ensemble model using SQM generated evidence. Performance is measured using multiclass classification metrics, including accuracy, macro-averaged precision, recall, and F1 score, complemented by qualitative assessment of reasoning clarity and action relevance. These experiments quantify the impact of SQM evidence enrichment and model ensembling on resolution accuracy, consistency, and analyst decision support quality.

\subsection{Evaluation Metrics}
Each module is evaluated using task-specific metrics reflecting both predictive accuracy and operational usefulness. 

\subsubsection{Threat Detection}
Threat detection is evaluated as a binary task using standard confusion matrix metrics: {Accuracy}, {Precision}, {Recall}, and {F1-score}. The {False Positive Rate (FPR)} is additionally reported due to its operational significance in SOC environments, where excessive false alerts directly increase analyst workload and contribute to alert fatigue:
\begin{equation}
\mathrm{FPR} = \frac{FP}{FP + TN}
\end{equation}
These metrics are applied consistently across both traditional ML models and LLMs.

\subsubsection{Query Generation}
Generated SIEM queries are evaluated for semantic fidelity using token-level similarity against manually constructed reference queries. {BLEU} score measures n-gram overlap:
\begin{equation}
\mathrm{BLEU} = \mathrm{BP} \cdot \exp \left( \sum_{n=1}^{N} w_n \log p_n \right)
\end{equation}
where $p_n$ is modified n-gram precision, $w_n$ are weighting factors, and BP is the brevity penalty. {ROUGE-L} captures longest common subsequence similarity to assess structural alignment:
\begin{equation}
\text{ROUGE-L} = \frac{(1 + \beta^{2}) \times \text{LCS}}{\text{Reference Length} + \beta^{2} \times \text{Prediction Length}}
\end{equation}
where LCS is the longest common subsequence length and $\beta$ controls recall-precision trade-off. Executable correctness is verified by direct execution on IBM QRadar and Google SecOps platforms. BLEU and ROUGE-L scores have demonstrated effectiveness in evaluating the semantic and structural quality of SIEM queries~\cite{saju2025synrag}.

\subsubsection{Threat Resolution and Recommendation}
Resolution code prediction is evaluated as a multiclass classification task using {Accuracy}, {Precision}, {Recall}, and {F1-score} (macro-averaged). For qualitative assessment of generated recommendations, an {LLM-as-a-Judge}\cite{gu2024survey} technique is employed, where a separate judge model scores each recommendation on a 0--10 scale across reasoning quality, action relevance, and overall usefulness:
\begin{equation}
\mathrm{Judge\ Score} = \frac{1}{N} \sum_{i=1}^{N} s_i
\end{equation}
The LLM-as-a-Judge scoring mechanism employs GPT-5.1 as an independent evaluator that was selected from outside the models evaluated in this paper for the purpose of unbiased scoring. The evaluation covers reasoning clarity, action relevance, and overall usefulness. This dual strategy captures both decision correctness and explanatory quality, both of which are essential for deployment in safety-critical environments.



\section{Results \& Analysis}

This section evaluates the proposed end-to-end threat management framework across its three core components: threat detection, evidence collection through query generation, and threat resolution with recommendation support. Results are reported using multiple quantitative metrics that cover predictive accuracy, operational reliability, and analyst-facing effectiveness. Comparisons are made against strong machine learning baselines and state-of-the-art large language models under controlled experimental settings.

\subsection{Threat Detection Performance}

Table \ref{tab:ml_llm_compact} reports detection performance across traditional ML models, individual LLMs, and the proposed ensemble. Among classical models, AdaBoost and XGBoost achieve the highest accuracy (0.671 and 0.672), but recall falls below 0.25 for both, meaning the majority of true critical threats go undetected. Their low FPR (under 0.05) shows that they use conservative decision boundaries, which are more suitable for precision-focused environments than for SOC operations, where false negatives can create serious operational risk. Logistic Regression (LogReg) and Decision Tree (DT) show a more balanced recall, close to 0.50, but they still have limited accuracy and F1 score. Naive Bayes (NB) performs poorly across all metrics because it cannot capture high-dimensional dependencies in enriched SIEM data. In contrast, LLMs perform much better than classical approaches in recall and F1 by using their strong ability to understand semantic patterns in both structured and unstructured log data. GPT-4o-mini achieves the best single-model balance: accuracy 0.826, recall 0.792, F1 0.782. DeepSeek-V3 and Qwen3-Next-80B attain high recall (0.848 and 0.958), but at the cost of FPR values of 0.445 and 0.776 respectively, making them operationally prohibitive due to alert overload.
The proposed majority-voting ensemble (GPT-4o-mini, Gemma-3n-E4B-it, Llama-3.3-70B) achieves the best overall tradeoff: accuracy 0.828, FPR 0.120, recall 0.708, F1 0.749. While recall is modestly lower than the most aggressive individual models, the substantial FPR reduction yields a more practical alerting profile means lower false positive rate and reduced fatigue for the analysts. This confirms that ensemble reasoning mitigates individual model biases and stabilizes decision boundaries for security-critical classification.

\begin{table*}[ht]
\centering

\begin{minipage}{0.60\textwidth}
\centering
\caption{Threat detection performance.}
\label{tab:ml_llm_compact}
\renewcommand{\arraystretch}{1.15}
\setlength{\tabcolsep}{4pt}
\small
\begin{tabular}{c l c c c c c}
\hline
 & Model & Acc & Prec & Rec & F1 & FPR \\
\hline
\multirow{5}{*}{\rotatebox{90}{ML}}
& LogReg & 0.590 & 0.484 & 0.551 & 0.515 & 0.384 \\
& DT & 0.608 & 0.504 & 0.492 & 0.498 & 0.316 \\
& AdaBoost & 0.671 & 0.789 & 0.230 & 0.357 & 0.040 \\
& XGBoost & 0.672 & 0.775 & 0.242 & 0.368 & 0.046 \\
& NB & 0.592 & 0.372 & 0.045 & 0.080 & 0.050 \\
\hline
\multirow{6}{*}{\rotatebox{90}{LLM}}
& gpt-4o-mini & 0.826 & 0.773 & 0.792 & 0.782 & 0.153 \\
& Qwen3 & 0.514 & 0.447 & 0.958 & 0.610 & 0.776 \\
& gemma & 0.701 & 0.598 & 0.744 & 0.663 & 0.327 \\
& Llama-3.3 & 0.760 & 0.672 & 0.770 & 0.717 & 0.246 \\
& DeepSeek-V3 & 0.671 & 0.555 & 0.848 & 0.671 & 0.445 \\
& gpt-4o & 0.621 & 0.514 & 0.758 & 0.613 & 0.469 \\
\hline
& Proposed & \textbf{0.828} & \textbf{0.795} & 0.708 & 0.749 & \textbf{0.120} \\
\hline
\end{tabular}
\end{minipage}
\hfill
\begin{minipage}{0.38\textwidth}
\centering
\caption{Query generation (BLEU, ROUGE-L).}
\label{tab:llm_bleu_rouge_compact}
\renewcommand{\arraystretch}{1.15}
\setlength{\tabcolsep}{5pt}
\small
\begin{tabular}{l c c}
\hline
Model & BLEU & ROUGE-L \\
\hline
Qwen3 & 0.149 & 0.522 \\
gpt-4o & 0.140 & 0.508 \\
DeepSeek-V3 & 0.137 & 0.492 \\
gemma & 0.121 & 0.495 \\
gpt-4o-mini & 0.091 & 0.462 \\
Llama-3.3 & 0.081 & 0.429 \\
\hline
\textbf{SQM} & \textbf{0.384} & \textbf{0.731} \\
\hline
\end{tabular}
\end{minipage}

\end{table*}

\subsection{Evidence Collection and Query Generation Evaluation}

\begin{figure}[t]
\centering
\begin{subfigure}[t]{0.45\columnwidth}
    \centering
    \includegraphics[height=3.8cm,keepaspectratio]{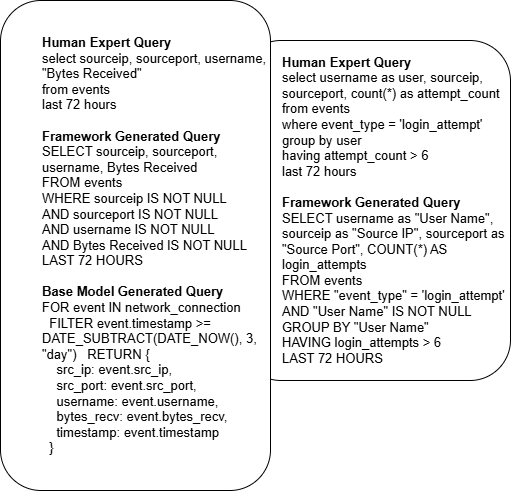}
    \caption{Framework vs. human and base LLM query.}
    \label{QueryComp}
\end{subfigure}
\hfill
\begin{subfigure}[t]{0.45\columnwidth}
    \centering
    \includegraphics[height=3.5cm,keepaspectratio]{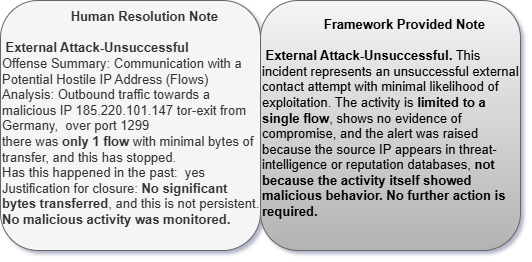}
    \caption{SOC analyst vs. LLM recommendation.}
    \label{RecommendCompare}
\end{subfigure}
\caption{Comparison of query generation and threat resolution.}
\end{figure}

Table  \ref{tab:llm_bleu_rouge_compact} evaluates the ability of large language models to generate SIEM executable queries under different settings using BLEU and ROUGE-L metrics. These metrics measure both lexical fidelity and structural overlap with reference queries, which are essential for executable correctness and semantic alignment. Base models without retrieval augmentation demonstrate limited performance. Even the strongest model, Qwen3-Next-80B-A3B-Instruct, achieves a BLEU score of only 0.149. This indicates that free form query generation struggles to produce correct syntax, operator usage, and field ordering required by SIEM query languages such as AQL and YARA-L. The variability across models further suggests that parameter scale alone does not guarantee syntactic reliability. The lack of explicit grammar and platform constraints limits the effectiveness of generic retrieval strategies for structured query generation. The proposed SQM architecture dramatically outperforms all baselines, achieving a BLEU score of 0.384 and a ROUGE-L score of 0.731. This represents more than a two-fold improvement over the strongest base model. Empirical evaluation identified GPT-4o as the most effective backbone model for the SQM architecture, producing the highest BLEU and ROUGE-L scores among all candidates. The gains can be directly attributed to three design factors. First, the syntax allow-lists enforce hard constraints on keyword usage and clause ordering, preventing invalid query constructions and hallucinating syntaxes. Second, metadata driven retrieval ensures that only semantically and operationally relevant exemplar queries are used as anchors. Third, documentation grounded prompting injects platform-specific rules that guide the model toward executable outputs. Together, these components transform query generation from an open-ended language task into a constrained synthesis problem, resulting in substantially higher correctness and consistency. We also evaluated the executable correctness of generated queries by testing them directly on both IBM QRadar and Google SecOps platforms. Without SQM, nearly all queries produced by the base models required some degree of manual modification before they could be successfully executed. In contrast, with the SQM architecture, 88\% of generated queries were syntactically correct and executed without any modifications, demonstrating the practical effectiveness of the proposed syntax-constrained metadata driven query generation approach. Figure \ref{QueryComp} compares ground-truth analyst queries with framework-generated outputs, showing that SQM achieves high structural and semantic alignment with expert-defined logic, but baseline LLM queries have syntax issues.



\subsection{Resolution Code Prediction with and without SQM}

Table 3 presents multiclass classification(Benign Positive, External Attack-Unsuccessful, External Attack-Successful, Insider Threat-Unsuccessful, Insider Threat-Successful, Escalated with No Response, and False Positive) results for resolution code prediction under different evidence settings. This task evaluates the downstream impact of SQM generated evidence on analyst decision support. Without SQM, performance varies significantly across models. The gpt-4o-mini model performs best in this setting with an accuracy of 0.783, indicating strong generalization from ticket text alone. However, other models show limited effectiveness. These results suggest that textual incident summaries alone are insufficient for consistent resolution prediction. Introducing SQM-generated evidence improves the performance of some models. The gemma-3n model shows the biggest improvement. In this model, the accuracy increases from 0.732 to 0.848 and F1 score rises from 0.698 to 0.834. This shows that structured, query-derived evidence provides useful signals that directly support resolution decisions. Models with little change, such as gpt-4o-mini, already perform near saturation even without extra evidence. We also created a weighted ensemble of Gemma-3n and GPT-4o-mini, enhanced with SQM-derived evidence, and it achieved the best overall resolution performance. It reached an accuracy and recall of 0.900, with a macro-averaged F1 score of 0.857. Model-specific weights of 0.60 for Gemma-3n and 0.40 for GPT-4o-mini were selected through grid search on the validation set to maximize accuracy.

\subsection{Recommendation Quality and Overall Analyst Support}

Table 4 reports the overall recommendation quality score, which aggregates multiple qualitative and quantitative dimensions, including correctness, clarity, relevance, and actionability. This score reflects the analysts' experience rather than isolated predictive accuracy. Across all baseline models, the inclusion of SQM consistently improves or maintained a stabilized overall score. This shows that evidence grounding improves not only classification performance but also the clarity and usefulness of the generated recommendations. The proposed system achieves the highest overall score of 8.70. It also outperforms the strongest baseline by a clear margin. Our proposed framework succeeds because it integrates a pipeline of detection, finding evidence, and resolution generation, in which each stage of processing enhance the subsequent stage. The system’s recommendations are generated based on the historical resolution data and analyst guidelines stored in the vector database. That's why our framework generated better recommendations and actions for the human analysts. Figure \ref{RecommendCompare} demonstrates strong alignment between the analyst-authored note and the framework-generated recommendation. Both classify the incident as External Attack–Unsuccessful and determine that no further action is required. The analyst’s observation of “only 1 flow with minimal bytes of transfer” and “No malicious activity was monitored” closely corresponds to the framework’s reasoning that the activity was “limited to a single flow” and “shows no evidence of compromise.” This overlap in findings and justification indicates high semantic and decision-level agreement between the automated system and human expert judgment.

\begin{table*}[t]
\centering
\setlength{\tabcolsep}{4pt}      
\renewcommand{\arraystretch}{1.15}  

\begin{minipage}[t]{0.58\textwidth}
\centering
\caption{Resolution code prediction performance.}
\label{tab:llm_classification_metrics}
\begin{tabular}{l l c c c c}
\hline
\textbf{Set} & \textbf{Model} & \textbf{Acc} & \textbf{Prec} & \textbf{Rec} & \textbf{F1} \\
\hline
\multirow{6}{*}{No SQM}
 & gpt-4o-mini   & 0.783 & 0.870 & 0.783 & 0.768 \\
 & Qwen3-Next-80B& 0.659 & 0.711 & 0.659 & 0.599 \\
 & gemma-3n      & 0.732 & 0.842 & 0.732 & 0.698 \\
 & Llama-3.3-70B & 0.565 & 0.667 & 0.565 & 0.527 \\
 & DeepSeek-V3   & 0.717 & 0.863 & 0.717 & 0.722 \\
 & gpt-4o        & 0.717 & 0.866 & 0.717 & 0.704 \\
\hline
\multirow{6}{*}{With SQM}
 & gpt-4o-mini   & 0.783 & 0.870 & 0.783 & 0.768 \\
 & Qwen3-Next-80B& 0.696 & 0.697 & 0.696 & 0.630 \\
 & gemma-3n      & 0.848 & 0.884 & 0.848 & 0.834 \\
 & Llama-3.3-70B & 0.600 & 0.789 & 0.600 & 0.578 \\
 & DeepSeek-V3   & 0.717 & 0.866 & 0.717 & 0.699 \\
 & gpt-4o        & 0.761 & 0.877 & 0.761 & 0.745 \\
\hline
Proposed & Ours & \textbf{0.900} & 0.825 & \textbf{0.900} & \textbf{0.857} \\
\hline
\end{tabular}
\end{minipage}
\hfill
\begin{minipage}[t]{0.38\textwidth}
\centering
\caption{Recommendation generation (LLM-Judge).}
\label{tab:llm_overall_scores}
\begin{tabular}{l l c}
\hline
\textbf{Set} & \textbf{Model} & \textbf{Score} \\
\hline
\multirow{6}{*}{No SQM}
 & gpt-4o-mini   & 8.61 \\
 & Qwen3-Next-80B& 8.24 \\
 & gemma-3n      & 7.91 \\
 & Llama-3.3-70B & 7.83 \\
 & DeepSeek-V3   & 8.54 \\
 & gpt-4o        & 8.52 \\
\hline
\multirow{6}{*}{With SQM}
 & gpt-4o-mini   & 8.63 \\
 & Qwen3-Next-80B& 8.41 \\
 & gemma-3n      & 8.43 \\
 & Llama-3.3-70B & 7.61 \\
 & DeepSeek-V3   & 8.59 \\
 & gpt-4o        & 8.57 \\
\hline
Proposed & Ours & \textbf{8.70} \\
\hline
\end{tabular}
\end{minipage}

\vspace{-3mm}
\end{table*}

Overall, the results demonstrate that the proposed framework consistently outperforms both traditional machine learning approaches and standalone large language models across all evaluated tasks. The ensemble detection strategy provides robust threat detection with reduced false positives. The SQM architecture enables reliable and executable query generation, overcoming a major limitation of generic LLM-based retrieval approaches, which is hallucinating syntax. Finally, the integration of SQM derived evidence into resolution and recommendation tasks yields substantial gains in accuracy, recall, and analyst-facing quality. Each incident ticket in ServiceNow includes a resolution time that covers the full processing time. The time includes investigation, evidence collection, resolution coding, and recommendation generation time. In the traditional manual workflow, the average resolution time was about 4 hours per ticket. With the proposed framework, the full processing time was reduced to about 10 minutes per incident. This shows a major reduction in analyst triage time while still maintaining high decision accuracy and recommendation quality. These results support the main idea of this work: constrained, metadata-driven, and ensemble-based LLM systems can perform effectively in real-world security operations and can do better than isolated models or unconstrained prompting alone.

\section{Conclusion}
This paper presents an end-to-end LLM-powered threat management framework designed to address the operational bottlenecks facing modern Security Operations Centers. By tightly coupling three components, namely ensemble-based threat detection, the SQM architecture for syntax-constrained query generation, and RAG-augmented resolution support. The framework automates the investigative workflow from alert triage through incident closure. The first component which is the ensemble detection module, achieves an accuracy of 82.8\% and a false positive rate of 0.120. This provides a good balance between detecting threats and reducing alert fatigue. In addition to classifying events as critical or non-critical, the system also assigns a risk score to each critical event. This score is based on SIEM-generated magnitude values and predicted criticality. It helps analysts focus on the highest-risk incidents first instead of treating all alerts the same. The second component, which we call SQM architecture, more than doubles baseline query generation performance with a BLEU score of 0.384 and ROUGE-L score of 0.731. The third component further enhances the resolution prediction performance by leveraging SQM derived evidence to increase accuracy from 78.3\% to 90.0\%. In addition, the second component is able to generate fully executable SIEM queries for 88\% of all inputs without human intervention. The recommendation quality score of 8.70 which reflects that analysts found the output as coherent and actionable. Overall, these gains reduce average triage time from approximately four hours to under ten minutes. The central finding of this work is that domain-constrained, retrieval-augmented LLM systems substantially outperform both traditional classifiers and unconstrained language models across every stage of the SOC workflow. Future directions include expanding SQM to additional SIEM platforms, integrating real-time threat intelligence, and establishing analyst feedback loops for continuous model refinement.








\appendix


\printbibliography[heading=subbibintoc]

\end{document}